\documentclass[twocolumn,secnumarabic,amssymb,amsmath,nofootinbib
]{revtex4}
\usepackage{graphicx}
\usepackage{epsf}
\usepackage{bm}
\usepackage{amsmath}
\usepackage{amsfonts}
\usepackage{amssymb}
\usepackage{epstopdf}
\usepackage{natbib}
\usepackage{rotating}
\usepackage{pdflscape}
\usepackage{color}
\usepackage{dcolumn}
\usepackage{ulem}
\usepackage{times}
\usepackage{appendix}
\usepackage{enumitem}
\usepackage{booktabs}
\usepackage{multirow}
\usepackage{orcidlink}
\begin{document}

\title{Non--relativistic Boson Stars as N--Body Quantum Systems}

\author{El\'ias Castellanos \orcidlink{0000-0002-7615-5004}}
\email{ecastellanos@mctp.mx} 
\affiliation {Mesoamerican Centre for Theoretical Physics\\ Universidad Aut\'onoma de Chiapas.
Ciudad Universitaria, Carretera Zapata Km. 4, Real del Bosque (Ter\'an), 29040, Tuxtla Guti\'errez, Chiapas, M\'exico.}

\author{Guillermo Chac\'{o}n--Acosta \orcidlink{0000-0002-7213-7386}}
\email{gchacon@cua.uam.mx} \affiliation{Departamento de
Matem\'aticas Aplicadas y Sistemas, \\Universidad Aut\'onoma
Metropolitana-Cuajimalpa, \\Vasco de Quiroga 4871, Ciudad de M\'exico, 05348, M\'exico.}

\author{Jorge Mastache \orcidlink{0000-0002-2709-6319}
}
\email{jhmastache@mctp.mx}
\affiliation{Consejo Nacional de Ciencia y Tecnolog\'ia, Av. Insurgentes Sur 1582, Col Cr\'edito Constructor, Del. Benito Ju\'arez, CP 03940. Mexico,}
\affiliation{Mesoamerican Centre for Theoretical Physics, Universidad Aut\'onoma de Chiapas, Carretera Zapata Km. 4, Real del Bosque (Ter\'an), 29040, Tuxtla Guti\'errez, Chiapas, M\'exico.}

\begin{abstract}
In the present work, we analyze the structural configuration of a collection of generic non--relativistic bosons forming a gravitational bound Bose--Einstein condensate that we interpreted as a non--relativistic boson star. We prove that the system's behavior can be obtained by analyzing its fundamental constituent's properties, i.e., the single particle properties.  Additionally, we show that by expressing the corresponding Newtonian gravitational potential, under certain circumstances, as a harmonic oscillator potential ones,  we can describe the conditions in which the non--relativistic boson star can form equilibrium configurations. In order to analyze the structural configuration related to the boson star, we employ four different \textit{ans\"atze} commonly used in the literature. The use of these \textit{ans\"atze} allows to compare the structural properties of the bosonic cloud or the boson star that leads to obtain several equilibrium configurations from compact objets matching to the size of typical stars to very gigantic systems comparable to the size of galaxy cluster dark matter halos. Finally, we show that these \textit{ans\"atze} predict, qualitatively speaking, the same structural and gravitational equilibrium configurations for different values of the parameters involved.
\end{abstract}

\maketitle

\section{Introduction}
\label{sec:intro}
Bose--Einstein condensates (BECs) play a very exciting and essential role in modern physics, relating a multitude of models spreading from microscopic well proved behavior of ultra--cold quantum gases to galactic and cosmological scales. Nevertheless, there is an issue not well understood in this scenario which deserves more in-depth study, i.e., the non--trivial conditions in which scalar fields can form BECs \cite{B1,B2,B3,B4,B5,B6,B7,Nos}. However, it seems to be that scalar fields in the form of some type of BEC formed by generic bosons can describe the basic properties of dark matter (DM) in the universe \cite{DM,DM1,DM2,DM3,DM4,DM5,DM6}. According to this line of thought, DM consists of a certain type of spin--zero bosons, such as ultra-light scalar field dark matter or fuzzy dark matter, weakly interacting massive particles, axions, etc.,
(depending on the specific model under consideration) which have not yet been observed.
The bosonic character of these particles, using the theory of relativistic Bose gases \cite{JB,LP}, 
also opens the door for the existence of scalar field dark matter in the 
form of BECs \cite{BoehmerHarko2007,ure}.\\
Complementary to the ideas mentioned above, some particular theoretical objects can be formed in a very similar manner. In some circumstances, a system of \textit{generic bosons} can form gravitationally bound  BECs leading to macroscopic objects,  
the so--called boson stars (BSs) \cite{1,2,3}.
On the one hand, some research lines suggest that BSs could be alternative candidates for black holes in the center of galaxies. On the other hand, it is generally accepted that the fundamental constituents of BSs are some generic scalars in the form of some type of BEC \cite{Eckehard}. This last assertion opens up the opportunity to describe these objects with BEC's formalism, which enriches the analysis.
We must mention that the study of BSs and its interpretation as BECs has been extensively analyzed \cite{SL}. Although these objects have also not been observed yet, their behavior and their structural properties lead us to think that these systems are highly related to scalar dark matter clouds in the universe. 
There is a \textit{zoo} of these objects (BSs) in the literature \cite{Eckehard}, basically characterized according to its dynamical behavior. For instance, BS could lie in the relativistic regime or not, 
they can also be characterized with respect to the type of self--interactions within the system \cite{Eckehard}, etc. 
The structural properties of the BS are quite interesting. For instance, Heisenberg's Uncertainty Principle provides pressure support in order to get a stable object. 
The size of the BS ranges from giant  to very compact objects
depending on the nature of the involved functional interactions among the constituents of the system. Finally, we must mention that the Klein--Gordon equation coupled with gravity describes the dynamics of BSs in the relativistic regime, and consequently, some Gross--Pitaevskii like--equations also coupled with gravity describe the dynamics in the non--relativistic limit, the so--called \textit{Gross--Pitaevskii--Poisson equation}, see for instance \cite{Kling1,Kling2} and references therein.
It is quite interesting that the non--relativistic limit of the self--interacting Klein--Gordon equation, which describes the dynamics of BSs, leads to the so--called Gross--Pitaevskii equation, when the scalar potential is of the form  $m^{2}|\phi|^{2}+\lambda |\phi|^{4}$, being $m$ the mass parameter and $\lambda$ the term related to self--interactions. Clearly, more general forms of the Gross--Pitaevskii equation can be used to generalize the corresponding interactions within the system. Those mentioned earlier lead to criteria of structural characterization of the BS, for instance, its size, stability, etc., by using (under certain conditions) the basic formalism behind usual laboratory BECs. 
 Strictly speaking, the Gross--Pitaevskii equation is an approximated equation valid for systems at zero temperature. Nevertheless, the predictions made by the Gross--Pitaevskii equation are a good approximation for temperatures $T <T_{c}$, where $T_{c}$ is the condensation temperature of the system. Such an equation can be used to analyze diluted weakly interacting systems' properties and when the number of particles is large enough for the condensed phase.
However, when the corresponding \textit{Gross--Pitaevskii--Poisson equation} is used to study the structural properties of the BS, the N--particles that constitute the system are analyzed as a single particle(--field). In other words, the nature of the phase transition provides the advantage to reduce the analysis of the $N$-body system to the analysis of the dynamics of a single body(--field) as in usual BECs. For this reason, we call the field appearing in the \textit{Gross--Pitaevskii--Poisson equation} the order parameter. The order parameter contains the information of the N--particles forming the condensed phase, and due to the highly correlated behavior of the BEC, this system behaves as a single entity. \\
In this work, we assume from the very beginning that the non--relativistic behavior of the BS can be described as a collection of weakly interacting generic bosons that are able to form gravitationally bound BECs. In other words, we describe some structural properties related to the BS through the quantum properties of its basic constituents, i.e., the properties of a single particle. \\
 The paper is organized as follows. In section \ref{sec:NB}, we study the fundamental properties of the BS viewed as a collection of bosons starting from the single particle description. We assume that the system behaves as a BEC. Also, we describe the approximation in which the Newtonian gravitational potential can be expressed as a harmonic oscillator potential that we interpreted as the trapping potential, like in usual laboratory BECs. In section \ref{sec:SBS}, we analyze the relevant structural functions that characterize the ground state of the system to obtain criteria of stability upon the BS. Finally, in section \ref{sec:CON}, we present a discussion, conclusions, and outlook. 
\section{N--Body Quantum System as a Non--Relativistic Boson Star }
\label{sec:NB}
As was mentioned in the introduction, the basic constituents of BSs are scalar particles(--field) (or spin zero--bosons), probably in the form of a BEC.
In this section we analyze the non--relativistic BSs behavior as a collection of bosons interpreted as a quantum $N$--body system,  in order to analyze some relevant properties associated with the bosonic cloud viewed as a BEC.
In this aim, we define the following $N$--body Hamiltonian which describes our non--relativistic BS
\begin{eqnarray}
 \label{MH}
 \hat{H}&=&-\frac{\hbar^{2}}{2m_{\phi}}\sum_{\delta,\gamma}\langle\delta\vert\nabla^{2}\vert \gamma\rangle\hat{a}^{\dag}_{\delta}\hat{a}_{\gamma}\\\nonumber&+&\frac{1}{2} \sum_{\delta,\gamma,\mu,\nu}\langle\delta,\gamma\vert V_{int} \vert\mu,\nu\rangle
 \hat{a}^{\dag}_{\delta}\hat{a}_{\gamma}\hat{a}^{\dag}_{\mu}\hat{a}_{\nu}\\\nonumber&+& \sum_{\delta,\gamma}\langle\delta \vert V_{g}\vert\gamma\rangle\hat{a}^{\dag}_{\delta}\hat{a}_{\gamma}
\end{eqnarray}
where $m_{\phi}$ is the mass of the boson particle and $V_{int}$ is the potential which describe the interactions within the system. Moreover, we have also inserted in Eq.\,(\ref{MH}) the contributions of the gravitational potential $V_{g}$. Additionally, the operators $\hat{a}$ and $\hat{a}^\dag $, correspond to the
creation and annihilation operators for bosons, satisfying the usual canonical commutation relations
\begin{equation}
[\hat{a}_{\mu},\hat{a}_{\nu}^{\dagger}]=\delta_{\mu \nu},\,\,\,\ [\hat{a}_{\mu},\hat{a}_{\nu}]=[\hat{a}_{\mu}^{\dagger},\hat{a}_{\nu}^{\dagger}]=0.
\end{equation}
As was mentioned above, the term, $V_{int}$ denotes the inter--particle potential, that will be assumed as $V_{int}\equiv U_{0}=\frac{4 \pi
\hbar^{2}}{m_{\phi}}a$, with $a$ the s--wave scattering length, i.e., at temperatures below the condensation temperature, only two--body interactions are taken into account. In other words, the system is diluted enough, and fulfills the condition $\rho |a|^{3}\ll 1$, where $\rho$ is the density of particles \cite{Dalfovo,Ueda,Pitaevski,Pethick}.
Additionally, $V_{g}$ depicts the contributions of the gravitational potential within the BS that we assume for simplicity with spherical symmetry.

Notice that in the corresponding $N$--body Hamiltonian
Eq.\,(\ref{MH}) we have the following terms
\begin{equation}
\label{ef1}
\langle\delta\vert{\nabla}^{2}\vert\gamma\rangle=\int{d}^{3}r{u}_{\delta}^{*}(\vec{r}){\nabla}^{2}{u}_{\gamma}(\vec{r}),
\end{equation}
\begin{eqnarray}
&& \langle\delta,\gamma\vert
V_{int}\vert\mu,\nu\rangle =\\\nonumber&&\int\int{d}^{3}{r}_{1}{d}^{3}{r}_{2}{u}_{\delta}^{*}(\vec{r}_{1}){u}_{\gamma}^{*}(\vec{r}_{2})
 U_{0}{u}_{\mu}(\vec{r}_{2}){u}_{\nu}(\vec{r}_{1}),
\end{eqnarray}
\begin{equation}
\langle\delta \vert
V_{g}\vert\gamma\rangle=\int{d}^{3}r{u}_{\delta}^{*}(\vec{r})V_{g}
{u}_{\gamma}(\vec{r}),
\end{equation}
where $\{{u}_{\gamma}(\vec{r})\}$ is a set of single--particle
functions.

Although the mean field solutions for the gravitational potential with no interactions, are known in terms of hypergemoeometric functions, one can in principle,
introduce some approximation to the gravitational potential. Let us consider a test particle in the outermost layer of the star, then its gravitational potential energy is proportional to the product of its masses, by the inverse of the distance to the center of the star that concentrates the largest part of the mass. For a spherically symmetric distribution, the first approximation is such that the mass is proportional to the central density times the volume of the sphere. Then, the gravitational potential would be proportional to the distance squared, i.e., a \textit{harmonic potential}. This approximation is made precise in Refs. \cite{Abel,Abel1}.

First, let us recall that the mass $M_{T}$ within a spherically symmetric BS, is given by
\begin{equation}
\label{eq:TMass}
M_{T}(r)=4 \pi\, m_{\phi} N \int_{0}^{r}\rho(r')\,{r'}^{2}\,dr'.
\end{equation}
where $0\leqslant r \leqslant R$. The radius $R$ will be considered the radius of the BS if most of its mass is contained in a region bounded by that radius. $N$ is the corresponding number of particles and $\rho(r)$ is the one particle probability density, which admits a series expansion when 
most of the matter of the BS is close enough to $r=0$, i.e.,
\begin{equation}
\label{eq:SE}
\rho(r)=\rho(r=0)+\sum_{n=1}^{\infty}\frac{\rho^{(n)}(r=0)r^{n}}{n\,!},
\end{equation}
where $m_{\phi} N\rho(r=0)$ is the central mass density of the BS, and $\rho^{(n)}$ its corresponding derivatives. 
However, the density $\rho(r)$ tends to infinity as $r \rightarrow \infty$ which is a non physical behavior since the density $\rho(r)$ must tends zero for large $r$. In order to avoid this unphysical scenario it is assumed that 
\begin{equation}
\label{assu}
\Bigg|\frac{\rho^{(n)}(r=0)}{\rho(r=0)}\Bigg| \approx \frac{1}{R^{n}}.
\end{equation}

Now, for the motion of a test particle that goes through the BS along the collinear diameter with the $z$-axis, the density peak is at the center of the BS, yielding
\begin{equation}
\label{TMr}
M_{T}(r)=\frac{4}{3}\pi\,m_{\phi} N \rho(r=0)r^{3}\Bigg[1+3\sum_{n=2}^{\infty}\frac{1}{(n+3) n\,!}\Bigl(\frac{r}{R}\Bigr)^{n}\Bigg].
\end{equation}
The sum in Eq.\,(\ref{TMr}) must be convergent, but small by hypothesis. Indeed, in Ref.\,\cite{Abel}, it is shown that $\forall \delta$  there is an index of the sum such that, for all the terms greater than that corresponding to such index, the sum is bounded by $ \delta $. For $\delta$ small enough but not negligible, the remaining terms of the sum can be considered as perturbations. Therefore, the main term of the potential is reduced to a harmonic potential with an \textit{effective gravitational frequency} given by 
\begin{equation}
\label{omegaG}
\omega_{g}=\sqrt{\alpha \pi Gm_{\phi} N \rho(r=0)},
\end{equation}
where $G$ is the Newtonian constant of gravitation. Here also $\alpha =4/3+4\delta$. If $\delta=10^{-1}$, corresponding to less than 10 \% of the total mass, the factor $ 52/ 30$ of Ref.\,\cite{Abel} is recovered. Briefly, we can define an \textit{effective gravitational frequency}, so that the gravitational potential can be interpreted at first order, as a trapping harmonic--like potential, as occurs in the usual BEC's formalism.
\bigskip

\section{Boson Star Structural Analysis}
\label{sec:SBS}
According to the conditions obtained in the previous section, the corresponding analysis for the properties related to the BS can be summarized as follows, the trapping potential is given by
\begin{equation}
\label{eq:potito}
V_{g}\approx \frac{1}{2}m_{\phi}\omega_{g}^{2}r^{2},
\end{equation}
and the approximation for the total mass
\begin{equation}
\label{totM}
M_{T}\approx 4 \pi\, N m_{\phi}\int_{0}^{R}\rho(r=0)\,r^{2}\,dr.
\end{equation}

On the other hand, if we further assume that most of the particles are inside the condensate, that is, in the $\vec{p}=0$ state then, this implies that the number of particles in the excited states is negligible for temperatures $T<T_{c}$, where $T_{c}$ is the condensation temperature. The contributions of the particles in the excited states could affect the properties of the system, see for instance Ref.\,\cite{Abel}. However, we consider here that almost all the particles lies in the corresponding ground state according to the approximation Eq.\,(\ref{eq:AP}). The contributions of the  excited states could be important in the stability analysis and will be analyzed in future works.
Thus, the last assertions can be expressed as follows.
\begin{equation}
\label{eq:AP}
N_{0}\approx N, \hspace{0.5cm}\sum_{\vec{p} \not=0} N_{\vec{p}} \ll N,
\end{equation}
being $N$ the total number of particles, $N_{\vec{p}}$ the number of particles in the excited states, and $N_{0}$ the number of particles in the ground state. Keeping terms up to second order in $\hat{a}_{0}$ and $\hat{a}^{\dag}_{0}$, i.e., $\langle \hat{a}_{0}^{\dag} \hat{a}_{0} \rangle=\langle N \rangle$, we are able to obtain the ground
state energy ($E_{0}$) associated with our BS
\begin{eqnarray}
 \label{E0}
 E_{0}&=&-\frac{{\hbar}^{2}}{2m_{\phi}}\langle0\vert\nabla^{2}\vert0\rangle N + \frac{1}{2}\langle0,0\vert U_{0} \vert0,0\rangle{N}^{2} \\ \nonumber
 &+& \langle0 \vert
 V_{g}\vert0\rangle N 
 \end{eqnarray}
Thus, we have for instance for the kinetic energy
\begin{eqnarray}
\label{Exa}
\langle0\vert\nabla^{2}\vert0\rangle=\int_{0}^{\infty}\int_{\Omega}^{}\Psi_{0}^{*}(r)\nabla^{2}\Psi_{0}(r){r}^{2}\sin\theta
dr d\theta d\phi,\,\,\,\, 
\end{eqnarray}
 and so on for each of the energy contributions to the ground state Eq.~(\ref{E0}).


\begin{widetext}

\begin{table}[h]
\centering
\makebox[\textwidth]{
\resizebox{0.9\columnwidth}{!}{\begingroup
\setlength{\tabcolsep}{10pt} 
\renewcommand{\arraystretch}{1.8} 
 \begin{tabular}{c c c c c c} 
 \hline \hline
  & & G  & E & LE & C \\ [1ex] 
 \hline
 $\Psi_{0}(r)$ & & ${\Bigl(\frac{{\beta}^{2}}{\pi}\Bigr)}^{3/4}{e}^{-{\beta}^{2}{r}^{2}/2}$ & ${\Bigl(\frac{{\beta}^{2}}{\pi^{2/3}}\Bigr)}^{3/4}{e}^{-{\beta}r}$ & ${\Bigl(\frac{{\beta}^{2}}{7^{2/3}\pi^{2}}\Bigr)}^{3/4} \left(1+r\beta \right){e}^{-{\beta} r}$ & $\sqrt{\frac{4\pi\beta^{3}}{(2\pi^{2}-15)}}\cos^{2}\left(\frac{\pi \beta r}{2}\right)$ \\ 
 A & & $\pi^{-3/2}$ & $\pi^{-1}$ & $\frac{1}{7\pi^3}$ & $\frac{4\pi}{(2\pi^{2}-15)}$ \\
 $\kappa$ & & $2.8$ & $4.2$ & $5.4$ & $1$ \\
 $\epsilon_1$ & & $\frac{3}{4}$ & $\frac{1}{2}$ & $\frac{3}{14\pi^2}$ & $\frac{(4\pi^{2}-6)\pi^{2}}{(24\pi^{2}-180)}$ \\
 $\epsilon_2$ & & $\frac{3}{4}$ & $\frac{1}{2}$ & $\frac{81}{28\pi^2}$ & $\frac{3(2\pi^{4}-5\pi^{2}+315)}{10(2\pi^{3}-15\pi)}$ \\
 $\epsilon_3$ & & $\frac{1}{2(2\pi)^{3/2}} $ & $\frac{1}{16 \pi}$ & $\frac{437}{25088\pi^5}$ & $\frac{35(24\pi^{3}-205\pi)}{288(2\pi^{2}-15)}$ \\ [1ex] 
 \hline \hline
 \end{tabular}
 \endgroup
 }}
 \caption{Ans\"atze table for the wave function of a single particle and its corresponding parameters $A, \kappa, \epsilon_1,\epsilon_3,\epsilon_3$. The ans\"atze presented here are known in the literature as the Gaussian (G), exponential (E), the linear exponential (LE) and the compact (C), respectively.}
\label{table:1}
\end{table}

\end{widetext}


At this point, we introduce some well--known \textit{ans\"atze} for the single-particle wave function $\Psi_{0}(r)$, which are summarized in the Table \ref{table:1}.
There are several ans\"atze used in the literature. As the wave functions for BS usually do not have compact support, three non--compact \textit{ans\"atze} are proposed, see for instance Ref.\,\cite{Joshua} and references therein. However, in the Thomas--Fermi approximation, the BS can have a fixed radius, which may have some advantages \cite{Joshua1}, for which we also propose a compact ansatz. Sometimes, these functions contain adjustable parameters to compare with the numerical solutions \cite{Joshua}. As we see in the Table \ref{table:1},  each of our proposed wave function has a single parameter $ \beta $ with units of the inverse of length. In principle, this parameter can be different in each case, but due to the approximation of the harmonic potential, we will assume that it fulfills ${\beta}=\sqrt{\frac{m_{\phi} \omega_g}{\hbar}}$. Moreover, as we will see later in the manuscript, its meaning is related to the BS's radius.
  
On the other hand, the probability density is the square of the wave function, $\rho(r) = |\Psi_{0}(r)|^{2}$. From this definition, we can obtain the corresponding central density evaluating at $ r = 0 $, i. e. 
 \begin{equation}
 \label{eq:r0}
 \rho(r=0)=   A\, \beta^3,
 \end{equation}
where $A$ is a numerical factor that depends on whether it is Gaussian (G), Exponential (E), Linear--Exponential (LE), or Compact (C) anzatz, according to table \ref{table:1}.
Notice that the central density in each case is expressed in terms of the parameter $\beta$. We can substitute these central densities in the expression for the \textit{effective gravitational frequency} Eq.\,(\ref{omegaG}) that gives us
   \begin{equation}
 \label{frecg00}
\omega_g =   \sqrt{\alpha \pi A G m_{\phi} N \beta^3 },
\end{equation}
also in terms of the inverse length $\beta$.

Let us realize that $\beta$ depends on the \textit{effective gravitational frequency}, and this, in turn, depends on the central density, i.e., by using Eq.\,(\ref{eq:r0}) we obtain
  \begin{eqnarray}\label{dens}
\rho(r=0)&=&   A \beta^3 = A \left( \frac{m_{\phi} \omega_g}{ \hbar} \right)^{3/2} \\\nonumber&=& A \left( \frac{m_{\phi} }{ \hbar} \right)^{3/2} \left(\alpha G\pi m_{\phi} N \rho(r=0)  \right)^{3/4}.
\end{eqnarray}
Therefore, if the \textit{ans\"atze} given in Table\,\ref{table:1} and the potential Eq. (\ref{eq:potito}) be compatible, then the following expression for the central density of the BS must be consistent
  \begin{equation}
\label{rocero10}
\rho(r=0)=  \left(\alpha G \pi  \frac{m_{\phi}^3}{\hbar^2} \right)^{3}A^4N^{3}.  
\end{equation}

Notice that for all practical purposes the \textit{effective gravitational frequency} for each ansatz has the same functional form, qualitatively speaking, and consequently the functional form of the central density for each ansatz has this shape. 

On the other hand, with the expression for $\beta$ in terms of the \textit{effective gravitational frequency}, if we substitute $\omega_g$ form Eq.\,(\ref{omegaG}), and central density from Eq.\,(\ref{rocero10}), we obtain an order of magnitude for the inverse length parameter $\beta$ for each ans\"atze, as follows
  \begin{equation}
  \label{beta000}
\beta = \frac{\alpha \pi G A N }{ \hbar^2}m_{\phi}^3.
\end{equation}

Although this characteristic length is not precisely the BS's radius, it gives us an estimate of its size. Indeed, the ratio between the BS radius and this length is a fixed quantity $ R\beta = \kappa$ that depends on the chosen ansatz \cite{Joshua}  as shown in table \ref{table:1} . In the case of the compact ansatz, the radius can be considered as the inverse of the $\beta$ parameter, $R = \beta^{-1}$.

Let us calculate the corresponding ground state energy of the BS,  by substituting each ans\"atze into the ground state energy Eq.\,(\ref{E0}), to obtain the following expression
\begin{eqnarray}
\label{eq:i}
 E_{0}=\epsilon_1\frac{{\hbar}^{2} \beta^2 N}{m_{\phi} } + \epsilon_2 \frac{m_{\phi}\omega_{g}^{2}N}{\beta^2} + \epsilon_3 U_{0}\beta^3 N^{2},
\end{eqnarray}
where the numerical coefficients $\epsilon_i$ differ for each ans\"atze and are also shown in Table  \ref{table:1}. Note that the radial integral in Eq.\,(\ref{E0}) cannot be performed up to infinity for the compact ansatz case, since one need to ask that the function vanishes for radii greater than $ \beta^{-1} $. In other circumstances, it is known that this can lead to some difficulties \cite{Joshua}. However, in this case, it is enough to integrate up to $ \beta^{-1} $ to obtain the numerical coefficients above.

To obtain the thermodynamic quantities, it will be necessary to replace $\beta$ and $\omega_g$ as functions of the volume. Since we are considering a spherically symmetric BS, the available volume in the ideal case,  i.e., when the interactions among the constituents within the BS are neglected, is as follows $V_{BS} = 4\pi \kappa^3\beta^{-3}/3$, so ground state energy becomes
\begin{eqnarray}
\label{eq:gii}
 E_{0} &=&\epsilon_1\frac{{\hbar}^{2}N }{m_{\phi} } \left(\frac{4 \pi \kappa^3}{3} \right)^{2/3} V_{BS}^{-2/3}    
 \\ \nonumber &+&  \epsilon_2  \pi A \alpha G  m_{\phi}^{2}N^{2}  \left(\frac{4 \pi \kappa^3}{3} \right)^{1/3} V_{BS}^{-1/3}  \\ \nonumber &+& \epsilon_3 U_{0}N^{2}  \left(\frac{4 \pi \kappa^3}{3} \right) V_{BS}^{-1},
\end{eqnarray}
 From the $N$--body ground state energy Eq.\,(\ref{eq:gii})  we therefore calculate the ground state pressure  $P_0 = -\frac{\partial E_0}{\partial V_{BS}}$ for each ans\"atze with the result 
\begin{eqnarray}
\label{eq:PI}
   P_{0} &=& \left( \frac{3}{4\pi \kappa^3}  \right)  \Bigl[  \frac{2}{3} \epsilon_1 \frac{\hbar^2 N}{ m_{\phi} }\beta^5  \\ \nonumber   &+& \frac{1}{3} \epsilon_2\pi A  \alpha G  m_{\phi}^2N^2  \beta^4 +  \epsilon_3 U_{0}N^{2} \beta^6\Bigr].
\end{eqnarray}

After rearranging the terms by identifying the scale $\beta$ from Eq.\,(\ref{beta000}), we can obtain the following two terms, which are of the same order in the length scale that those usually found, but with different coefficients 
\begin{eqnarray}
\label{eq:P2}
   P_{0} &=&   \left( \frac{3}{4\pi \kappa^3}  \right)  \left[ \frac{\left( 2\epsilon_1 + \epsilon_2 \right)}{3}    \frac{\hbar^2 N}{m_{\phi} }\beta^5   +   \epsilon_3 U_{0}N^{2} \beta^6  \right]. 
\end{eqnarray}
%

\begin{figure}[h!]
    \includegraphics[width=0.39\textwidth]{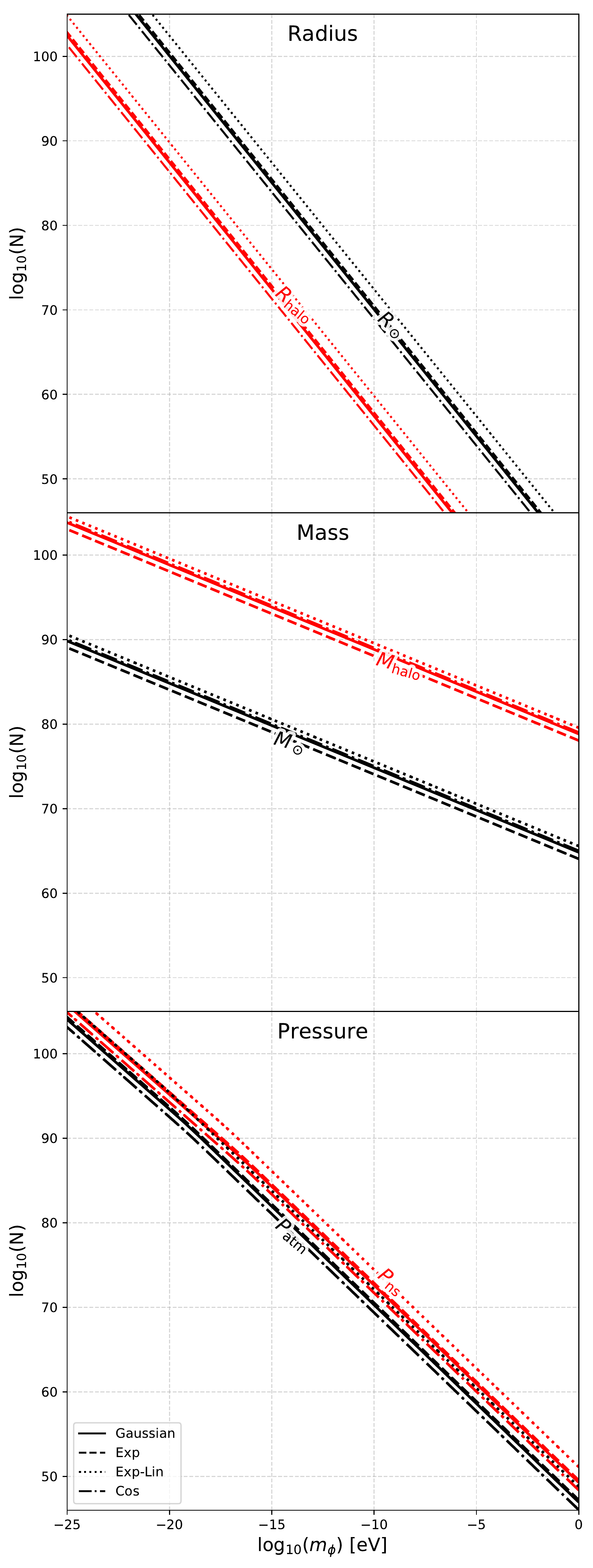}
    \caption{\footnotesize{ Plots of the Radii (top panel), Mass (middle panel) and Pressure (Bottom panel) for a Boson star. Red (and black) lines represent extreme high (low) values for each case, for the radii panel are a dark matter halo characteristic radius of $10^{14}$ Mpc, $R_{\rm halo}$ (in red), and the Solar radius, $R_\odot$ in black.  In the case of the mass, we show contour lines for a solar mass, $M_\odot$ (in black), and a typical mass for dark matter halos $10^{12} \, M_\odot$ , $M_{\rm halo}$ (in red). For the pressure we show lines that represent the pressure of a neutron star $\mathcal{O} (10^{31})$ Pa, $P_{\rm ns}$ (in red), and 1 atmosphere of pressure, $P_{\rm atm}$ (in black). Different line--styles represent the different approximations, straight, dashed, dotted, dash--dotted represent the gaussian, linear--exponential, exponential and cosine, respectively.}}
    \label{fig:eos} 
\end{figure}

If we assume that the pressure and gravity allows the BS to remain in equilibrium, then the following constraint to the number of particles is reached $N_e$,
\begin{equation}
	N_e = \frac{\left(2\epsilon_1 + \epsilon_2 \right) }{12\pi \kappa \epsilon_3}\frac{R}{|a|},
\end{equation} \label{eq:scat_len}
where clearly $N$ is several orders of magnitud greater than $\kappa$. The scattering length, whose value can also be negative, should be only constrained from the particle physics model, this is, from Eq.\,\eqref{eq:scat_len} given a value for $a$ we should know the region where systems are not allowed to exist due to equilibrium. Clearly, if $N < N_e$ gravity overcome pressure and we could have an implosion of the system, if $N > N_e$ then pressure overcome gravity and apparently the system becomes unstable. $N_e$ allow us to find the systems that are in equilibrium, and stability condition will be study in a future work, where the rotation of the BS can be also included. 

Alternatively to this scenario a phenomenological stability condition has been proposed in Ref.\,\cite{cornell} for a trapped laboratory BEC. For a system with an attractive interaction, there is not enough kinetic energy to stabilize the BEC and it is expected to implode. A BEC can avoid implosion only as long as the number of atoms is less than a critical value given by
\begin{equation}
\label{eq:NC}
N_{c}=\gamma ^{2}\frac{R}{|a|},
\end{equation}
%
where the parameter $\gamma^{2}$ is the so--called stability coefficient and $R$ the size of the system. The the stability coefficient depends on the properties of the trapping potential, see Ref.\,\cite{cornell} for details. Thus, according to our model  the stability coefficient is in each case
%
	
\begin{equation}
	 \gamma^{2} = \frac{\left(2\epsilon_1 + \epsilon_2 \right) }{12\pi \kappa \epsilon_3}.
\end{equation}\label{eq:NC1}
Introducing Eq.\,\eqref{eq:NC} into Eq.\,\eqref{eq:P2}, we notice that the scattering length, $a$, must have positive sign, otherwise both terms in Eq.\,\eqref{eq:P2} are identical but opposite in sign. If $a>0$ the pressure simplifies as follows
%
\begin{equation}
	P_e =  \frac{\left(2\epsilon_1 + \epsilon_2 \right)^2}{24 \pi^2 \epsilon_3}  \frac{\kappa \hbar^2}{a m_\phi R_e^4},
\end{equation} \label{eq:P3}
where the subindex $e$ means that is evaluated at gravitational equilibrium, $N = N_e$.

To compute the equilibrium number of particles of the system, $N_e$, we still need to solve Eq.\,\eqref{eq:NC} because $R$ is function of the number of particles, $N$. Using $R = \kappa/\beta$ and Eq.\,\eqref{beta000} we have, 
\begin{equation}
\label{eq:n_equilibrium}
	 N_e  = \left( \frac{(2\epsilon_1 + \epsilon_2)\hbar^2}{ 12\pi^2 A \epsilon_3  a G m_\phi^3 \alpha }  \right)^{1/2}.
\end{equation}

then, the only free parameter is the scattering length, $a$. 

 \begin{table*}[t]
\centering
	\begin{tabular}{c||l|llll}
	& & Gaussian & Exponential & Lin-Exp & Compact \\
	\hline
	\multirow{4}{*}{ Sun--like } & $m_\phi$ [eV] & $ 1.29 \times 10^{-12} $ & $ 5.15 \times 10^{-12} $ & $ 8.51 \times 10^{-12} $ & $ 2.92 \times 10^{-13} $ \\
	& N & $ 5.23 \times 10^{76} $ & $ 2.19 \times 10^{75} $ & $ 4.31 \times 10^{76} $ & $ 3.44 \times 10^{77} $ \\
	& a [m] & $ 9.6 \times 10^{-69} $ & $ 3.96 \times 10^{-67} $ & $ 6.35 \times 10^{-66} $ & $ 2.47 \times 10^{-63} $ \\
	& P [Pa] & $ 3.03 \times 10^{12} $ & $ 1.17 \times 10^{11} $ & $ 2.66 \times 10^{12} $ & $ 2.75 \times 10^{20} $ \\
	\hline
	\multirow{4}{*}{ Dwarf halo } & $m_\phi$ [eV] & $ 1.35 \times 10^{-22} $ & $ 5.37 \times 10^{-22} $ & $ 8.87 \times 10^{-22} $ & $ 3.04 \times 10^{-23} $ \\
	& N & $ 5.02 \times 10^{94} $ & $ 2.1 \times 10^{93} $ & $ 4.14 \times 10^{94} $ & $ 3.3 \times 10^{95} $ \\
	& a [m] & $ 9.21 \times 10^{-75} $ & $ 3.8 \times 10^{-73} $ & $ 6.09 \times 10^{-72} $ & $ 2.37 \times 10^{-69} $ \\
	& P [Pa] & $ 4.22 \times 10^{-20} $ & $ 1.62 \times 10^{-21} $ & $ 3.7 \times 10^{-20} $ & $ 3.83 \times 10^{-12} $ \\
	\hline
	\multirow{4}{*}{ Cluster halo } & $m_\phi$ [eV] & $ 2.01 \times 10^{-26} $ & $ 8.02 \times 10^{-26} $ & $ 1.33 \times 10^{-25} $ & $ 4.54 \times 10^{-27} $ \\
	& N & $ 3.36 \times 10^{104} $ & $ 1.41 \times 10^{103} $ & $ 2.77 \times 10^{104} $ & $ 2.21 \times 10^{105} $ \\
	& a [m] & $ 6.17 \times 10^{-83} $ & $ 2.54 \times 10^{-81} $ & $ 4.08 \times 10^{-80} $ & $ 1.59 \times 10^{-77} $ \\
	& P [Pa] & $ 1.05 \times 10^{-14} $ & $ 4.03 \times 10^{-16} $ & $ 9.17 \times 10^{-15} $ & $ 9.5 \times 10^{-7} $ \\
	\end{tabular}

  \caption{ Values to form a Sun--like or galaxy cluster halo system for the different ans\"atze the gaussian, the exponential, the linear exponential (Lin--Exp) and the compact. $m_\phi$ is the mass of the particle, $N$ is the number of particles, $a$ is the scattering length, and $P$ is the pressure of the system.}
\label{table:2}
\end{table*}

\section{Numerical analysis}
\begin{figure}[t]
    \includegraphics[width=0.45\textwidth]{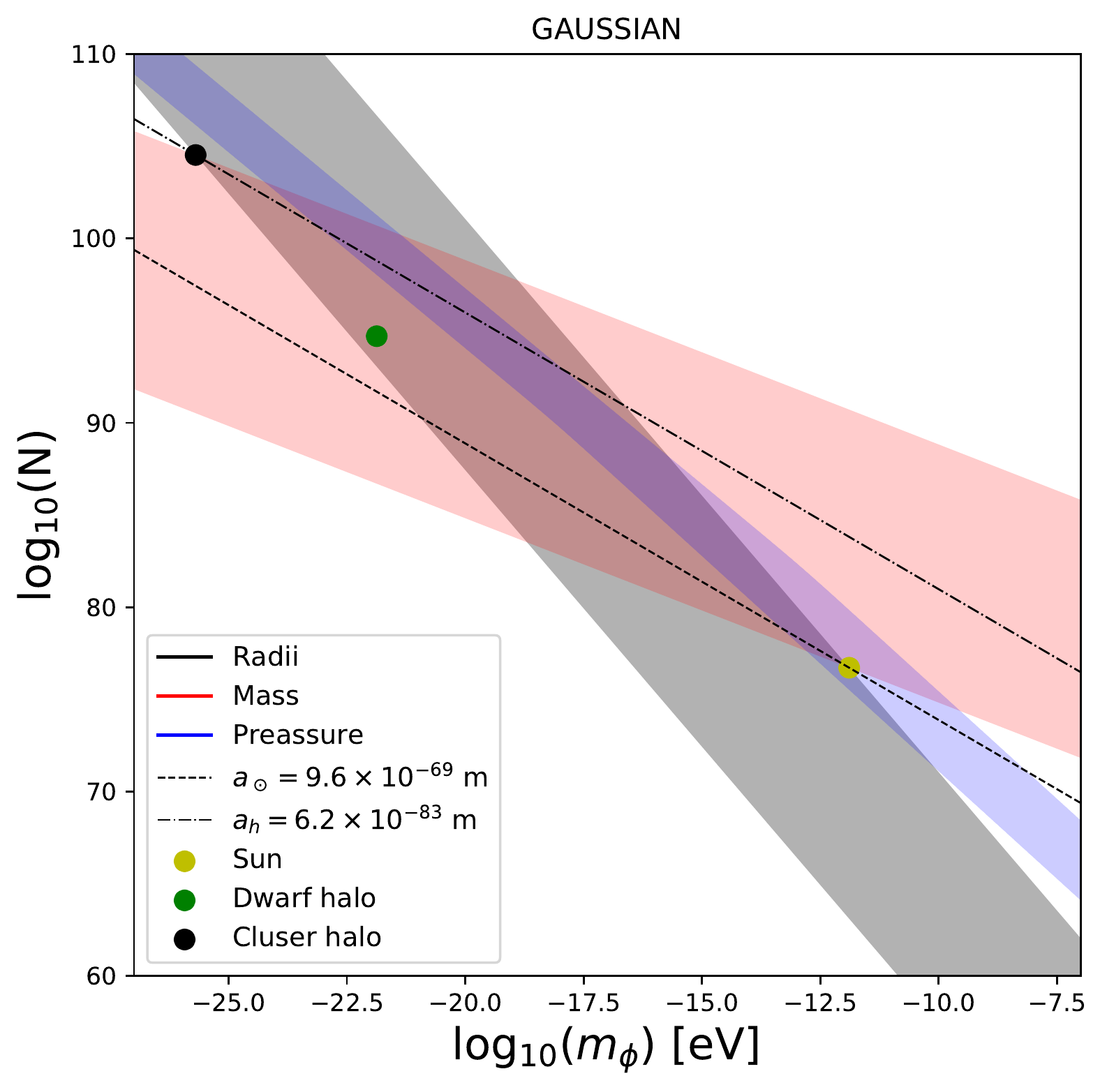}
    \caption{\footnotesize{Contour plots where shaded regions may represent realistic astrophysical systems. Black region represents system of the size between $R_\odot$ and a typical dark matter halo in a galaxy cluster $10^3$ kpc. Red region represent system between a solar mass ($M_\odot$) and the mass of a dark matter halo in a galaxy cluster ($10^{14} \; M_{\odot}$). Blue region represent systems between 1 atmosphere and inner crust neutron star ($10^{31}$ Pa) pressure. Black lines are system in equilibrium given by Eq.\eqref{eq:n_equilibrium}, dashed (dash-dotted) line is taken $a$ for a Sun--like (dark matter halo kind) system. Yellow, green, and black dots represent the Sun, dwarf DM halo, and galaxy cluster dark matter halo system, respectively.}}
    \label{fig:all_gausian} 
\end{figure}

In Fig. \ref{fig:all_gausian}, black lines represent systems in equilibrium, in all the figures we have taken two extreme examples. The dashed line is defined using $a_{\odot}$ is for a system of with a mass ($M_e = M_\odot$) and radius ($R_e = R_\odot$) as the Sun. Dash--dotted lines is for a system defined with the value of $a_h$ to represent a system of the size and mass of a typical cluster of dark matter halo, this is $M_{e} = 10^{14} \; {\rm M_\odot}$ and $R_e = 10^3$ kpc \cite{Newman:2012nv}.

We find interesting that with only one parameter this approach can predict BECs of very different scales, and may be consider a fine--tuning problem to differentiate the value of the scattering length $a$. The larger value for $a$ is give to describe Sun--like systems and the compact hypothesis gives its largest values, which is of the order of $10^{-63}$ m. For each of the hypothesis taken, the values of $a$ is different  depending if one wants to describe a Sun--like, dwarf DM halo, or a cluster DM halo system, see the Appendix \ref{appendix1} in which we show all the cases. It is interesting that, independent of the hypothesis taken, the value of $a_{sun}$ is 14 order of magnitud bigger than the case for the cluster dark matter halo, $a_h$. 

Additionally, by using Eqns.\,\eqref{eq:gii}, \eqref{eq:P2}, and \eqref{eq:n_equilibrium} we can find a relation between the pressure, $P$, and the energy density, $\varrho$, to compute the equation of state (EoS) of the type $P  = \omega\varrho$. Thus, we obtain
\begin{equation}\label{eq:w_eos}
 	\omega = \frac{1 + \left( \frac{N}{N_e} \right)^2}{\frac{3(\epsilon_1 + \epsilon_2)}{2\epsilon_1 + \epsilon_2} + \left( \frac{N}{N_e} \right)^2} \;\; .
\end{equation}
When $N \gg N_e$ then $\omega \sim 1$ this may describe systems with properties know as a \textit{stiff matter}, see for instance Ref.\,\cite{chavanis} and references therein for details on this kind of schemes. For systems in equilibrium, Eq.\,\eqref{eq:w_eos} simplifies
\begin{equation}
	\omega =\frac{1 + \frac{2\epsilon_1}{\epsilon_2}}{2\left(  1 + \frac{5\epsilon_1}{4\epsilon_2}  \right)},
\end{equation}
then, for the gaussian and exponential case we have $\epsilon_1/\epsilon_2 = 1$, this is, $\omega = 2/3$ describing an ideal system. For the exponential case $\epsilon_1 / \epsilon_2 = 2/27$ and $\omega = 31/59 \approx 0.53$ and the compact case $\omega \approx 0.63$. Notice that for systems in equilibrium, the equation of state does not depend on any of the variables for the different models, and this also coincide with the fact that the scattering length, $a$, is very small, which reflect the nature of an ideal non--relativistic gas. 

Despite the value of $\omega$ in the EoS, the fact that the pressure of the system, Eq.\,\eqref{eq:P3}, depends on $a$ make possible to form halos in equilibrium of a galaxy cluster size with very small pressure, this characteristics resemble the properties of the dark matter. For instance, to form a system with the properties a dark matter halo that soround a galaxy cluster in the gaussian case we would need a scattering length of $a_h = 6.17\times 10^{-83}$ m, a boson particle of mass $m_\phi = 2.01\times10^{-26}$ eV, and $N = 3.36\times10^{104}$ particles, and the BEC would have a pressure of $1.05\times10^{-14}$ Pa, it is the extension of the condensate that makes it plausible to existence of this kind of systems.

To form system with mass and size of the Sun, for instance, we would need $N = 5.23 \times 10^{76}$ particles of mass $m_\phi = 1.29\times 10^{-12}$ eV with an scattering length $a_{sun} = 9.6\times 10^{-69}$ m. This Sun--like systems would have a pressure  of $P_e = 3.03 \times 10^12$ Pa. The shadowed areas In Fig.\,\ref{fig:all_gausian} represent systems with reasonable radii, mass and pressure for an astrophysical system. It is also interesting that systems with very high pressure can be found in this scheme, for instance, systems with $N = 9.3 \times 10^{91}$ particles with mass $m_{\phi} = 8.0\times 10^{-18}$ eV and scattering length $a = 9.6 \times 10^{-84}$ m would have a pressure of $P = 1.5\times 10^{31}$ Pa, which is of the order of the inner crust pressure of a neutron star \cite{1603.02698}. However the approach we are proposing may not be valid to such a high pressures and relativistic corrections may have to be taken into account. Notice also that  when quantum effects are taken into account, the pressure tends to zero at zero temperature for bosons \cite{pathria}. Conversely, the equation of state $P=\varrho_{e}$ in which the pressure is proportional to the energy density $\rho_{e}$ is known in the literature as the so--called \textit{stiff matter}, see for instance Ref.\,\cite{chavanis} and references therein. We must mention that we are able to obtain apparently a \textit{stiff matter} equation of state at $T=0$, even when our BS lies in the non--relativistic and low--density regime. The topics mentioned above deserves more in--depth analysis and will be presented elsewhere \cite{jaja}.

Finally, several previous works \cite{Padilla:2020jdj,Matos:2000ss,Matos:2000ki,Lora:2011yc,RodriguezMontoya:2010zza,Harko:2011zt,Hu:2000ke} have put constrains on the mass of scalar fields  using galaxy rotation curves of dwarf galaxies since it is believe that the kinematic of this kind of systems is dominated by a DM halo. A typical size of the DM halo for a dwarf galaxy is of the order 22 kpc and have a total mass of the order of $10^8 \; {\rm M_\odot}$  \cite{Kravtsov:2012jn, Walker:2009zp}. To form systems with this features we would need a boson particle with a mass of $m_\phi = 1.35 \times 10^{-22}$ eV and a scattering length $a = 9.21 \times 10^{-75}$ m, this is consistent within an order of magnitude with previous results, and it is consistent with the DM as dust hypothesis since the pressure is very small $P = 4.22 \times 10^{-20}$ Pa.

A relevant difference between anzats is the order of magnitude for the pressure of the systems in the compact scenario i.e., the compact ansatz, in which case is up to 9 orders of magnitude bigger than the exponential ansatz for the cluster halo and the Sun--like examples. It is seem to be that this discrepancy in the pressure between the compact and the non--compact ans\"atze relies on the choice for the corresponding BS's radius $R=\beta^{-1}$, i.e., $\kappa=1$ for the compact ansatz, see Eq.\,(\ref{eq:P3}). In other words, the results obtained in the present work agree with previous results reported in Ref.\,\cite{Joshua}, which suggests that the case of the compact ansatz deserves deeper study and should be handled carefully.

\section{Conclusions}
\label{sec:CON}

We have analyzed a collection of non--relativistic gravitational bounded generic bosons forming a Bose--Einstein condensate starting from the single particle properties. We have also proved that the system can form stable structures in several scenarios that can be interpreted as BSs. By using four \textit{ans\"atze}, the Gaussian, Exponential, the Linear exponential (non--compact ansatz), and the Cosine (compact ansatz), we are able to prove that they predict almost the same structural configuration, qualitatively speaking. Additionally,  we have shown that different values of the corresponding scattering length, together with some specific values of the corresponding number of particles, lead to several sizes of BS in gravitational equilibrium. With our model, we can obtain from compact objects (i.e., the size of the sun, for instance) to gigantic configurations comparable to the size of galaxy cluster dark matter halos. In other words, our model predicts several configurations that may be stable and can form systems in gravitational equilibrium in a wide range of sizes. 
Notice that we are able to extract also significant properties associated with the BS thermodynamics. For instance, concerning the equation of state, we have calculated the corresponding ground state energy for each ansatz, from which we can obtain the corresponding internal energy and, consequently, the corresponding pressure. We can define two apparent limits according to the definition of the pressure $P=-\partial E_{0}/\partial V_{BS}$. The first one corresponds to the ideal case, i.e., $U_{0}=0$, and the second one at zero temperature. For the first case, we obtain that the equation of state is given by 
$P=\frac{2}{3}\frac{E_{0}}{V_{BS}}=\frac{2}{3} \varrho_{e}$. Conversely, at zero temperature (in which we can neglect the contributions of the kinetic energy) the equation of state can be expressed approximately as $P=\frac{E_{0}}{V_{BS}}=\varrho_{e}$.
Let us remark that the aforementioned equations of state are irrespective of the considered ansatz. Notice that $P=\frac{2}{3}\varrho_{e}$ is the standard equation of state for an ideal non--relativistic bosonic system (it can be proved also that this is the equation of state for a system of ideal non--relativistic fermions in the classical regime \cite{pathria}). Additionally, when quantum effects are taken into account, the pressure also tends to zero at zero temperature for bosons \cite{pathria}. Conversely, the equation of state $P=\varrho_{e}$ in which the pressure is proportional to the energy density $\varrho_{e}$ is known in the literature as the so--called \textit{stiff matter}. 
We must mention that we are able to obtain apparently a \textit{stiff matter} equation of state at $T=0$, even when our BS lies in the non--relativistic and low--density regime. The topics mentioned above deserves more in--depth analysis and will be presented elsewhere \cite{jaja}.
Finally, the present work must be extended to rotating systems in order to analyze the corresponding stability. Moreover, more general interactions within the system could be also relevant for compact objects, i.e., three--body interactions, and so on. Consequently, logarithmic interactions within a compact non--relativistic BS could be able to describe these scenarios. One more issue related to the BS configurations described in the present work is that they can be useful, in principle, to estimate the quantity of DM in the solar system. The topics mentioned above also deserve more in--depth analysis to study the eventual relation of BSs and DM as BECs of generic bosons in the universe.
\acknowledgments
E.C. acknowledges the receipt of
the grant from the Abdus Salam International Centre for Theoretical Physics, Trieste, Italy. This work was supported also by CONACyT M\'exico under Grant No. 304001.


\appendix 
\section {Plots}
\label{appendix1}
Here we present the contour plots for the exponential, linear-exponential and compact cases.

\begin{figure}[h!]
    \includegraphics[width=0.45\textwidth]{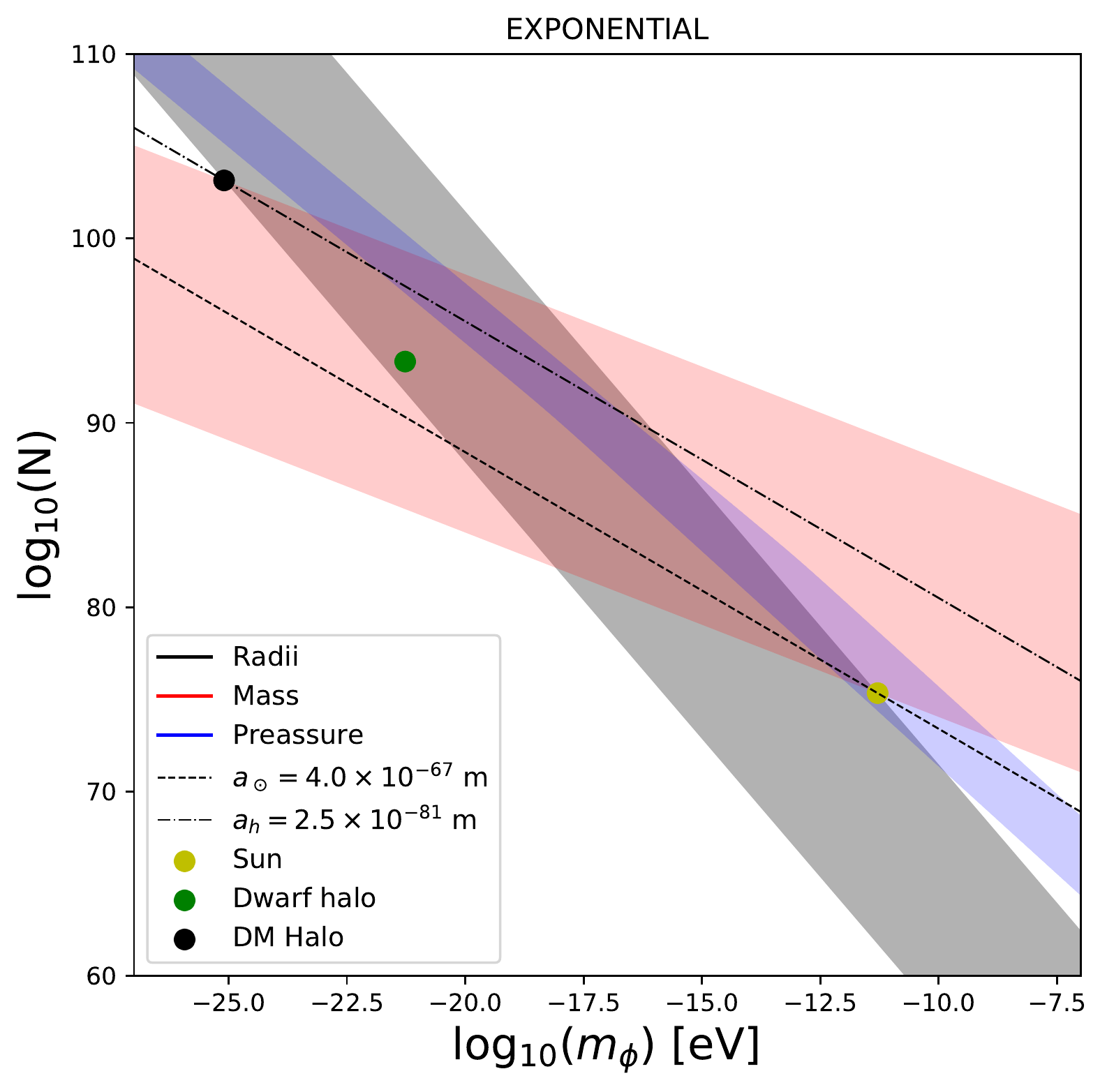}
    \caption{\footnotesize{ Contour plots for the exponential ansatz. Colors and styles are the same as in Fig.\ref{fig:all_gausian}.  }}
    \label{fig:all_exp} 
\end{figure}

\begin{figure}[h!]
    \includegraphics[width=0.45\textwidth]{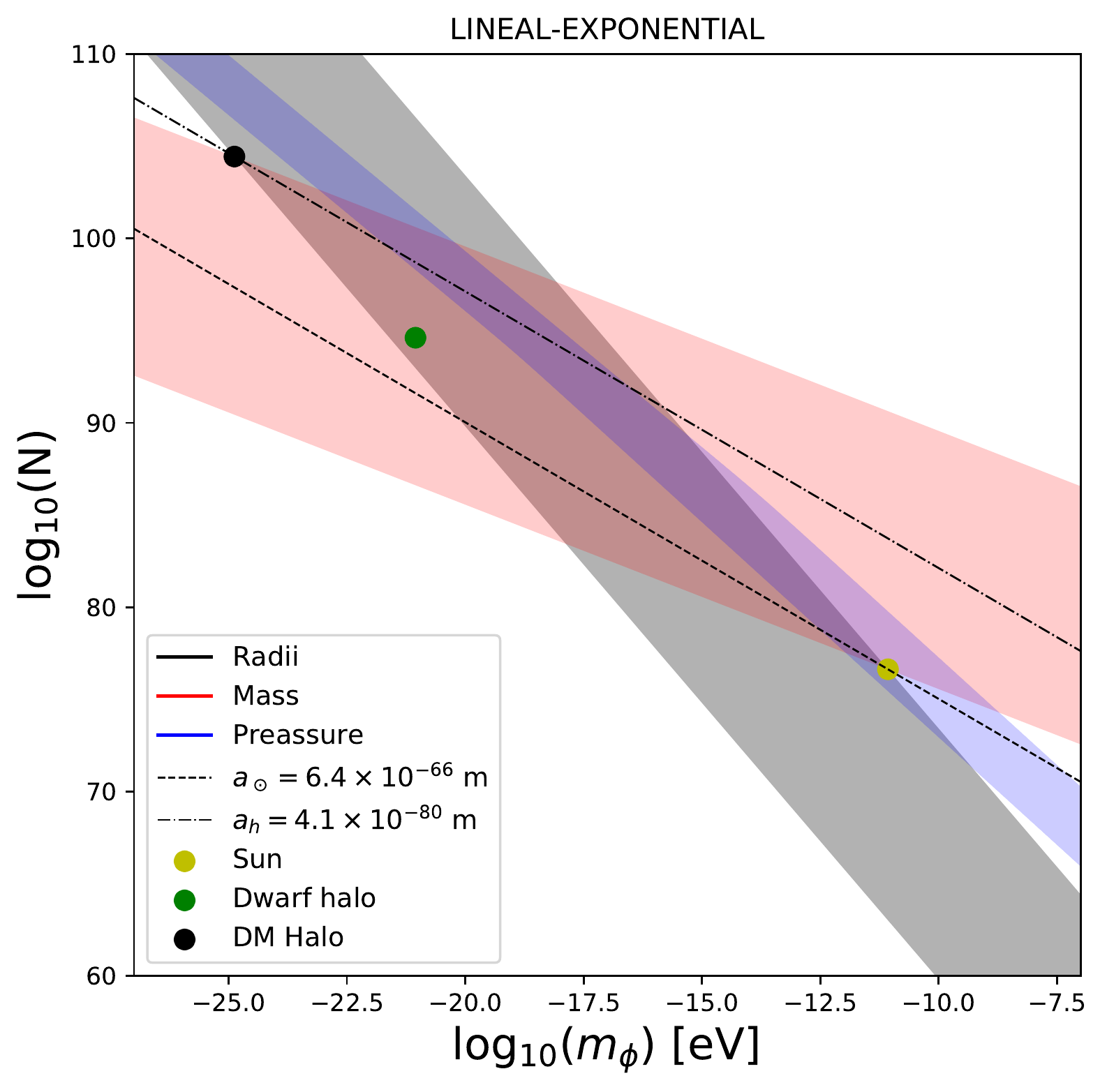}
    \caption{\footnotesize{  Contour plots for the linear-exponential ansatz. Colors and styles are the same as in Fig.\ref{fig:all_gausian}.}}
    \label{fig:all_linexp} 
\end{figure}

\begin{figure}[h!]
    \includegraphics[width=0.45\textwidth]{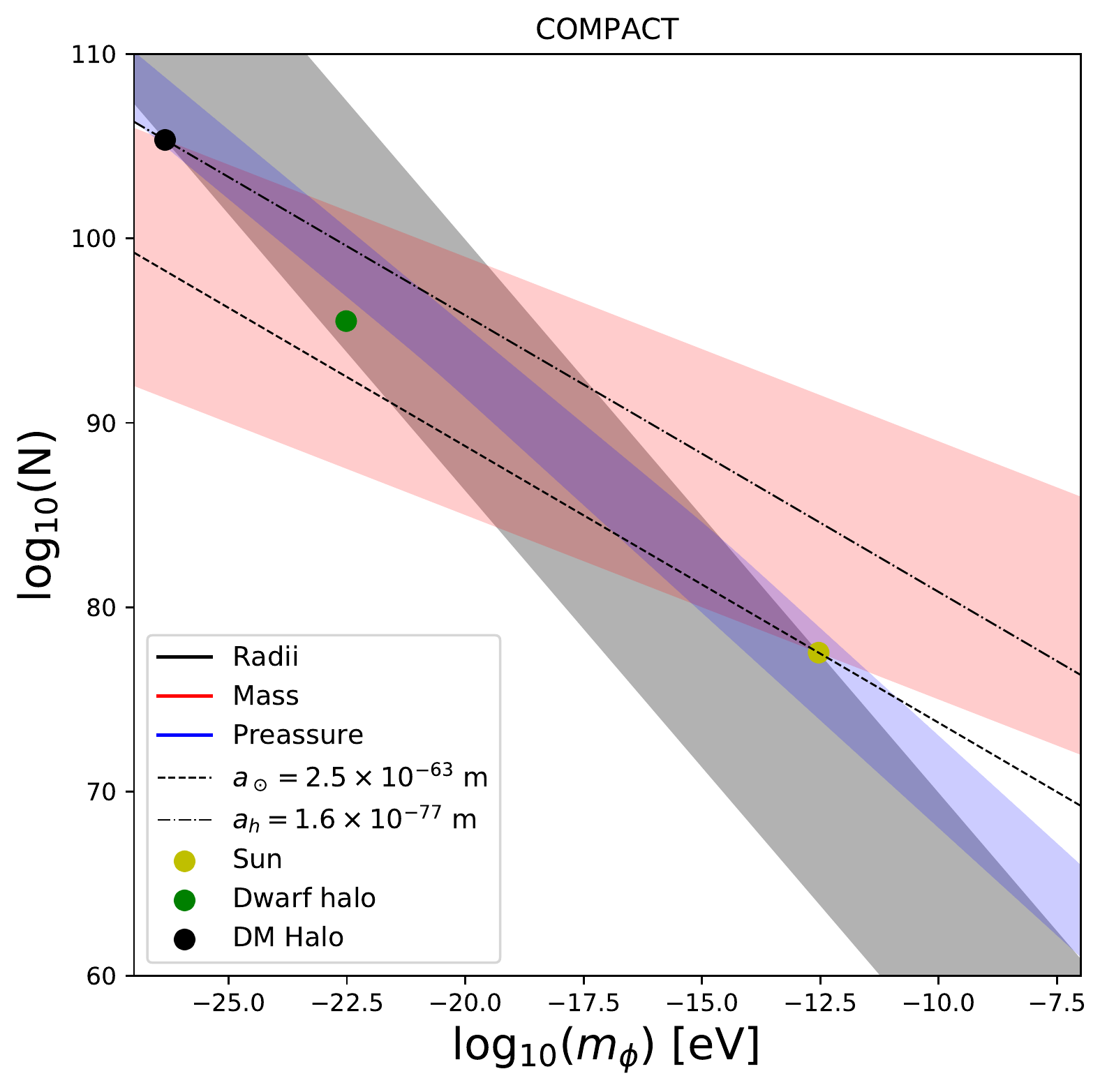}
    \caption{\footnotesize{ Contour plots for the compact ansatz. Colors and styles are the same as in Fig.\ref{fig:all_gausian}. }}
    \label{ffig:all_cosine} 
\end{figure}


\end{document}